\documentclass[12pt]{article}
\usepackage{graphics}

\begin{document}

\textheight 8.5in
\textwidth 6.5in
\topmargin -30pt

\def\double{\baselineskip 18pt \lineskip 10pt}
\setcounter{page}{0}
\thispagestyle{empty}

\renewcommand{\arraystretch}{1.5}

\def\be{\begin{equation}}
\def\ee{\end{equation}}
\def\ba{\begin{eqnarray}}
\def\ea{\end{eqnarray}}
\def\bq{\begin{quote}}
\def\eq{\end{quote}}
\def\PL{{ \it Phys. Lett.} }
\def\PRL{{\it Phys. Rev. Lett.} }
\def\NP{{\it Nucl. Phys.} }
\def\PR{{\it Phys. Rev.} }
\def\MPL{{\it Mod. Phys. Lett.} }
\def\IJMP{{\it Int. J. Mod .Phys.} }
\newcommand{\labell}[1]{\label{#1}\qquad_{#1}} 
\newcommand{\labels}[1]{\vskip-2ex$_{#1}$\label{#1}} 
\newcommand\gapp{\mathrel{\raise.3ex\hbox{$>$}\mkern-14mu
              \lower0.6ex\hbox{$\sim$}}}
\newcommand\gsim{\gapp}
\newcommand\gtsim{\gapp}
\newcommand\lapp{\mathrel{\raise.3ex\hbox{$<$}\mkern-14mu
              \lower0.6ex\hbox{$\sim$}}}
\newcommand\lsim{\lapp}
\newcommand\ltsim{\lapp}
\newcommand\M{{\cal M}}
\newcommand\order{{\cal O}}

\newcommand\extra{{\rm {extra}}}
\newcommand\FRW{{\rm {FRW}}}
\newcommand\brm{{\rm {b}}}
\newcommand\ord{{\rm {ord}}}
\newcommand\Pl{{\rm {pl}}}
\newcommand\Mpl{M_{\rm {pl}}}
\newcommand\mgap{m_{\rm {gap}}}
\newcommand\gB{g^{\left(\rm \small B\right)}}

\begin{center}
{\large{\bf Large Extra Dimensions and Cosmological Problems}}

\vspace*{0.3in}
Glenn D. Starkman$^1$, Dejan Stojkovic$^1$ and
Mark Trodden$^2$
\vspace*{0.3in}

\it
$^1$ Department of Physics \\
Case Western Reserve University \\
Cleveland, OH 44106-7079, USA \\
\vspace*{0.2in}

$^2$ Department of Physics \\
Syracuse University \\
Syracuse, NY, 13244-1130 USA \\
\vspace*{0.2in}

\end{center}
\vfill

\begin{abstract}
We consider a variant of the brane-world model in which the universe is 
the direct product of a Friedmann, Robertson-Walker (FRW) space
and a compact hyperbolic manifold of dimension $d\geq2$.
Cosmology in this space is particularly interesting.
The dynamical evolution of the space-time leads to the injection of
a large entropy into the observable (FRW) universe.
The exponential dependence of surface area on distance in hyperbolic geometry
makes this initial entropy very large,
even if the CHM has relatively small diameter (in fundamental units).
This provides an attractive reformulation of the cosmological entropy problem, 
in which the large entropy is a consequence of the topology,
though we would argue that a  final solution of the entropy problem
requires a dynamical explanation of the topology of spacetime.
Nevertheless, it is reassuring that this entropy can be
achieved within the holographic limit if the ordinary FRW space is also
a compact hyperbolic manifold. 
In addition, the very large statistical
averaging inherent in the collapse of the initial entropy onto the brane 
acts to smooth out initial inhomogeneities. This smoothing is then sufficient 
to account for the current homogeneity of the universe.
With only mild fine-tuning,  the current flatness of the universe can also then
be understood.
Finally, recent brane-world approaches to
the hierarchy problem can be readily realized within this framework.
\end{abstract}

\vfill

\noindent CWRU-P12-00 \\

\noindent SU-GP-00/12-1 \hfill hep-th/????????

\eject

\baselineskip 17pt plus 2pt minus 2pt

\section{Introduction}
\label{Introduction}
The standard initial value problems of Big Bang cosmology
are usually addressed by a period of cosmic inflation \cite{inflation}, to which
viable alternatives have been elusive. (See however \cite{MAD,varyingc}.)
In this paper, we will consider the status of these problems in the
context of the idea that
our $3+1$ dimensional universe is only a submanifold (3-brane)
on which Standard Model fields are confined inside
a higher dimensional space \cite{Anton,ADD,RS,otherearly}.
In such models, only gravitons and other geometric degrees of freedom
may propagate in the bulk space-time.
An important motivation for these theories
has been to solve the so-called hierarchy problem ---
explaining the largeness of the Planck mass compared to the scales
characterizing other interactions, in particular to the weak scale.
If $M_{F}$ is the actual fundamental scale of gravitational interactions,
and if the volume of the extra dimensional d-manifold is $V_\extra$, 
then \cite{Anton,ADD,RS,otherearly} by Gauss's law,
at distances larger than the inverse mass of the lightest Kaluza Klein (KK) mode in the 
theory,
the gravitational force will follow an inverse square law with an effective
coupling of 
\be
\label{Vextra}
\Mpl^{-2} = M_{F}^{-(d+2)}V_\extra^{-1}
\ .
\ee
The canonical realization of this scenario \cite{ADD} assumed
that the extra-dimensional manifold, $\M_\extra$,
was a $d$-torus of large spatial extent (in fundamental units, $M_{F}^{-1}$).
In that case, $M_{F}\gsim 50$TeV is consistent with existing particle physics
and cosmological phenomenology for $d\geq2$.
However, from many points of view $d$-tori are special.
Because they admit flat (Euclidean) geometries, they have no intrinsic geometric scale,
and so there is no {\it a priori} reason that they should have such a large extent.
There is also no gap in the graviton spectrum to the first KK mode.
Further, compact manifolds which admit a flat geometry are a set of measure zero
in the space of $d$-manifolds.

Recently, it was argued \cite{KMST} that if the extra dimensions comprised
a compact hyperbolic manifold (CHM) then the same volume suppression
of gravity could be obtained, and all constraints avoided with a manifold
whose radius is only $\order(30) M_{F}^{-1}$.
In this case $M_{F} \gsim 1$TeV is  allowed
and the observed gauge hierarchy is a consequence of the topology of space.
For $d=2$ and $3$, most manifolds are compact hyperbolic.
(For $d>3$, there is no complete classification of compact manifolds; indeed,
most compact manifolds probably admit no homogeneous geometry.)
These manifolds can be obtained from their better known universal covering space $H^d$
by ``modding-out'' by a (freely-acting) discrete subgroup $\Gamma$ of the isometry
 group of $H^d$.
(Just as $d$-tori are obtained by modding out d-dimensional Euclidean space $E^d$ by 
a freely-acting discrete subgroup of the
Galilean group in $d$-dimensions.) If the structure of the full manifold is
\be
\Sigma_{d+4} = {\bf R}\times {\cal M}_{FRW}\times {\cal M}_{extra} \ ,
\ee
then the metric on such a space can be written as
\be
ds^2= g^{(4)}_{\mu\nu}(x)dx^{\mu}dx^\nu + R_c^2 g^{(d)}_{ij}(y)
dy^{i}dy^j.
\label{metric}
\ee
Here $R_c$ is the physical curvature radius of the CHM,
and $g_{ij}(y)$ is therefore the metric on the CHM normalized so that
its Ricci scalar is ${\cal R}=-1$.

Clearly, unlike Euclidean geometry, hyperbolic geometry has an intrinsic scale ---
the radius of curvature $R_c$.
One therefore expects that if there is a gap in the graviton spectrum,
then
\be
\mgap = {\cal O}(R_c^{-1}) \ .
\ee
In $d=2$ and $3$ (and probably in $d>3$ as well), there is a countable infinity of  CHMs
with volumes distributed approximately uniformly from a finite minimum value to infinity
(in units of $R_c^d$). An important property of hyperbolic geometry is that
at large distances
volume grows exponentially with radius. As an example, note that in $H^d$,
the volume internal to a $d-1$-sphere of radius $L\gg R_c$ is given by
\be
\label{VofL}
V(L)\sim R_c^d e^{\beta} \ ,
\ee
where
\be
\beta \equiv \left[\frac{(d-1) L}{R_c}\right] \ .
\ee
The volume of a CHM is therefore  of the same form (\ref{VofL}), where $\beta$
is a constant determined by topology and related to the maximum spatial extent of the 
manifold.
$e^\beta$ is a measure of the manifold's ``complexity'';
in $d=2$, it is proportional to the  Euler characteristic of the  manifold.

While primarily motivated by attractive particle physics features,
this construction admits a host of interesting cosmological possibilities.
In this letter we address the entropy, flatness and homogeneity problems
in the context of models with compact hyperbolic extra dimensions.
We argue that in this particular class of spatial manifolds,
with generic particle-physics content,
well-motivated initial conditions lead to observable universes that are old,
flat and homogeneous, like our own.

\section{Cosmological Problems and Extra Dimensions}
To understand how this picture works, let us first review why
inflation is so successful. During an inflationary
epoch \cite{inflation} the universe expands superluminally by a large factor,
meanwhile supercooling and ``storing'' energy  in the inflaton
field. After inflation, decay of the inflaton field results in
the release of this stored energy into relativistic particles and
an enormous increase in the
total entropy of the universe. This leaves the entropy density of
the universe everywhere much higher than if the universe had
cooled adiabatically while undergoing a standard FRW expansion by
the same factor. As a result of this expansion and entropy
production, the large-scale homogeneity, flatness and
entropy problems of cosmology are resolved (For a discussion see, 
for example, \cite{Lin90}). 
It is not possible to obtain
such a result from ordinary subluminal expansion, since this would
require maintaining a constant entropy density through the expansion, in
violation of the $3+1$ dimensional Einstein equations.

\subsection{The Entropy Problem}
One alternative attempt to solve at least some of the cosmological
problems involves traditional Kaluza-Klein theories
\cite{alvarez}-\cite{abbott}. The idea is that the universe possesses extra
spatial dimensions beyond the three that we observe. Some of
these extra dimensions may be contracting while our $3$
dimensions are expanding. In this process, entropy could be
squeezed out of the contracting extra dimensions,
filling the three expanding ones, although it remains to understand the existence 
of the large total entropy in the universe.

In the model of \cite{kolb} the metric is taken to be
\be
ds^2=-dt^2 + a(t)^2 dx_{\mu} dx^{\mu} +b(t)^2 dy_i dy^i \ ,
\ee
with $\mu = 0,\ldots,3$, $i= 1,\ldots,d$.
Both scale factors
($a$ for the ordinary and $b$ for the extra space) are dynamical, and the
evolution begins from zero volume, i.e. an
initial singularity. The scale factor of the extra space reaches a maximum
value and recollapses to a final singularity. As $b$ approaches the
final singularity, $a$ goes to
infinity. This dynamics is such that the total volume,
\be
V_{\rm TOT} \propto a^3 b^d \equiv \sigma^{d+3} \ ,
\ee
actually decreases,
i.e. the extra scale factor is decreasing more rapidly than the ordinary
one is increasing. (Here, $\sigma$ is the geometric mean scale factor.)

It should be mentioned that the
classical equations are not to be believed all the way back to
the initial time $t_0=0$, where quantum effects could change the whole picture. 
Therefore, one imagines starting from some finite time,
perhaps at an energy
scale close to the fundamental energy scale of $M_{F}$.
Also, for a realistic theory some
stabilization mechanism for the extra dimensions is required. This mechanism
would prevent $b$ from becoming arbitrarily small. Since the fundamental physics
is governed by the scale of $M_{F}$, it seems reasonable to expect $b$ to
stabilize close to $M_{F}^{-1}$.
Nevertheless, despite these considerations, a careful analysis of
such cosmologies \cite{kolb} shows that there is
insufficient $3+1$ dimensional entropy production in these models
to solve the entropy problems.

However, in the context of large extra dimensions, the very large volume of the
extra dimensions is a source of much greater entropy than in traditional
Kaluza-Klein theories. Also, there is a new
effect. Entropy will continue to fall onto the brane even after
the stabilization of the extra dimensions. The massive
gravitational modes (Kaluza-Klein excitations), which are
nevertheless massless from the $4+d$ point of view, cannot decay
into two other massless particles if the extra-dimensional
momentum is conserved. This implies that the massive gravitons
can live for a very long time since they can not decay into the
empty bulk. However, the presence of the brane breaks the
translational invariance and allows momentum non-conservation in
the extra dimensions if decay takes place on the brane. The decay
of these modes would be preferentially to Standard Model
particles propagating on the brane, or to these plus the graviton zero mode,
which is just the ordinary $4$-dimensional graviton.
The coupling of gravitational
modes to non-gravitational modes is typically unsuppressed
compared to the coupling to other gravitational modes. Since
there are many light non-gravitational modes on the brane, and
only one light gravitational mode on the brane, we expect most
decays of bulk gravitational modes to deposit their entropy in
Standard Model fields.
Finally, given that $M_{F}\geq$TeV, the universe will (as shown below)
thermalize before nucleosynthesis  and then
evolve normally after the end of the entropy condensation era.

\begin{figure}[!ht] \label{fig}
\centering
\includegraphics{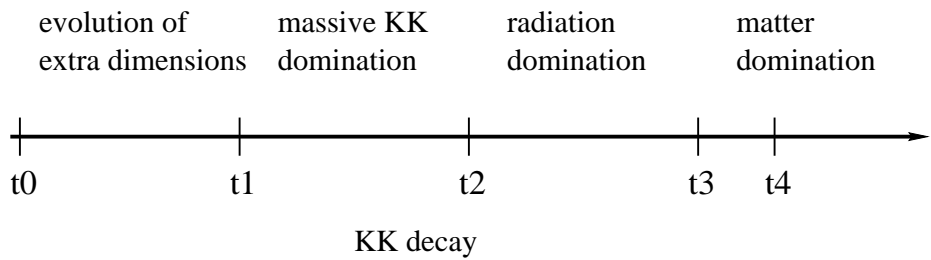}
\caption{Dynamical evolution of a universe with large extra dimensions}  
\end{figure}

We will consider this phenomenon in detail, in the specific case that both
the extra dimensions and our FRW space are described by CHMs. Thus, we will write 
their respective volumes as
\begin{eqnarray} \label{V}
V_{\rm FRW} & = & a(t)^3 e^{\alpha} \nonumber \\
V_{\rm extra} & = & b(t)^d e^{\beta} \ ,
\end{eqnarray}
where $a(t)$ and $b(t)$ are the respective curvature scales, and $\alpha$ and $\beta$ are
topological constants.
To understand this better, let us divide the evolution of the universe
into several different eras. We imagine that the universe appears, in the sense that
it's geometry can first be treated classically, at time $t_0$. 
The era of dynamical evolution of the ordinary and extra spaces begins at $t_0$ and ends at
$t_1$ when the extra dimensions are stabilized. After
stabilization, the era during which massive KK modes dominate
follows from $t_1$ to $t_2$, at which point the massive KK modes decay, 
and the entropy moves from the bulk to the brane. 
This leads into the usual radiation dominated era,
from $t_2$ to $t_3$, and the matter dominated era, from $t_3$ to
$t_4$ (see Fig. \ref{fig}).

It has been argued that
the maximum entropy which can be contained in a space is equal
to the area of the boundary of the space in fundamental units.
This is the Holographic Principle \cite{holography}, which
we may use to provide a bound on the total initial entropy of the universe.
The Holographic Principle asserts that the total entropy grows not
as the volume but as the area of the extra-dimensional manifold.
The attractive feature of $\M_\extra$ and $\M_\FRW$ being
of the form $H^n/\Gamma$ is that the area of a large CHM is approximately
equal to the volume (up to geometric constants of order unity) in fundamental units.
Further, the area of a CHM, like its volume, grows exponentially with radius.
Thus the maximal total initial entropy is
\be
S_0^{max} = g_0 a_0^3 e^\alpha b_0^d e^\beta T_0^{d+3} \ ,
\ee
where $g_i (i=0,1,2)$ is the density of states for a given
era. This is also a reasonable estimate for the actual value of $S_0$, and
we will take $S_0 = S_0^{max}$. Requiring $S_0$ to be
at least $10^{88}$ (which is the entropy within the horizon today),
not taking the contribution from $g_0$ into account, we infer
\be
e^{\alpha + \beta} \gsim 10^{88} 
\ee
if $a_0^{-1} \sim b_0^{-1} \sim T_0 \sim M_{F} \sim$TeV.
Relation (\ref{Vextra}) implies that $\beta$,
a topological constant of the extra space, should not be larger than 80. 
However, there is no such constraint on $\alpha$, 
and so the large initial entropy 
can be achieved within the holographic limit with $\alpha \gsim 130$. 
(Alternatively, one could drop the assumption that the FRW sub-manifold
is hyperbolic, and this becomes merely a lower limit on its total volume.

This brings us to the cosmological entropy problem --
the need for the enormous dimensionless number $10^{88}$) to specify the state of 
our universe
\cite{Lin90}.
We see that while we have not arrived at a dynamical solution of the problem,
we have found that the problematically large dimensionless number may reasonably be regarded
as the exponential of a much smaller dimensionless number -- the ratio of the 
size of the small compact dimensions of spacetime to their curvature scale.
One may of course ask why $\alpha$ and $\beta$ should be so large. 
Since there are a countable infinity of two- and three-
dimensional CHM-s with arbitrarily large volume, a better question might
perhaps be why $\alpha$ and $\beta$ are so small. 
Sidestepping the possible anthropic
arguments, we suggest that such questions can only truly be answered
in the context of a quantum cosmology incorporating dynamical topology change.

The total entropy in the universe for $t_1<t<t_2$ is
\be
S_1 = g_1 a_1^3 b_1^d e^{\alpha + \beta} T_1^{3+d}  \ ,
\ee
where $a_i$, $b_i$ denote $a(t_i)$ and $b(t_i)$ respectively.
Now, as $t \to t_2$ the universe approaches a temperature $T_*$, at which the
massive KK modes decay. During this decay, entropy is not conserved, and so
we must estimate the temperature on the
brane after decay. To do this, note that energy density is conserved during the
decay, and make the approximation that the decay is instantaneous, so that
at $t=t_2$ the temperature undergoes a rapid change from $T_*$ to $T_2$,
after which the
universe ceases to be matter dominated (since the massive KK modes responsible for
this have now decayed). Equating the energy densities at $t_2$, before and after
the decay of the KK modes yields
\be
\rho_*\equiv g_* T_*^{3} M_{KK}= g_2 T_2^{4} \equiv \rho_2 \ ,
\ee
where $g_* T_*^{3}$ measures the number density of KK modes at
temperature $T_*$. Thus,
\be
T_* = \left(\frac{g_2}{g_*}\right)^{1/3}
\left(\frac{T_2^4}{M_{KK}}\right)^{1/3}  \ .
\ee
We must now require that $T_2 \geq 1$MeV, when the usual radiation-dominated era begins,
so that the results of standard big bang nucleosynthesis are not changed. 

The total entropy of the universe after the decay of KK modes is
\be
S_2 = g_2 \frac{M_{KK}}{T_2} S_*
\ee
where $S_*$ is the total entropy just before the decay
\be
S_* = g_* a_*^3 b_*^d e^{\alpha + \beta} T_*^{3+d} \ .
\ee
Here, $a_* = a_2$ because the decay is instantaneous and $b_*=b_1$ because
the extra dimensions are stabilized after $t=t_1$.
An  aspect of the entropy problem closely related to the homogeneity
problem is that the entropy contained within the horizon at early times 
(for example recombination or nucleosynthesis) was much less than the present 
value of about $10^{88}$. This means that the present Hubble volume consisted
of many causally disconnected regions at some earlier time and causal
processes could not produce the smooth universe we see today.
We must therefore require not only that the total
entropy in the universe be at least $10^{88}$, but also that the entropy within the
horizon, at the beginning of the ordinary evolution, i.e. $t=t_2$, is $10^{88}$.
The horizon volume in CHM geometry, i.e. a 3-dimensional sphere of 
radius $L(t)$, cut out of an $H^3$ of curvature radius $a(t)$ is
\be \label{VL}
V_L=\pi a(t)^3 \left[ \sinh \left( \frac{2L(t)}{a(t)} \right) 
-\frac{2L(t)}{a(t)} \right]
\ee
where the horizon radius $L$ is given by
\be
L(t) = a(t) \int \frac{dt}{a(t)} \ ,
\ee
while $a(t) = a_0 \left( \frac{t}{t_0} \right)^p$. The coefficient $p$ depends
on dynamics and it is unknown for $t_0<t<t_1$ and $p=2/3$ for $t_1<t<t_2$. 
At the moment $t=t_2$, practically the whole entropy has already
fallen onto the brane. Thus, if by then the 3-horizon volume $V_L$ contains 
a curvature volume $V_{\rm FRW}$ within it, the entropy problem is solved
because $V_{\rm FRW}$ already contains entropy of $10^{88}$.
{}From (\ref{V}) and (\ref{VL}) we see that this requires $t_2/t_0$
to be greater than $10^3$ and $10^4$ for $p=1/2$ and $p=2/3$ respectively,
which can be easily satisfied with any reasonable dynamics. For example, 
one possible choice $T_1 \sim$TeV and $T_2 \sim 100$MeV, which  also
explains the flatness and homogeneity of the universe (see subsections (\ref{Flatness}) and 
(\ref{Homogeneity})), gives $t_2/t_1 \sim 10^6$.

\subsection{The Flatness Problem}
\label{Flatness}
Let us now turn to the flatness problem -- the fact that observations today
show no trace of a curvature of the universe although Einstein's equations dictate that even 
a small initial curvature term quickly dominates over matter or radiation density
in the evolution of the universe.

Expressed in fundamental units, the initial energy density of the
universe is
\be
G_{4+d}\rho_0=\frac{g_0}{M_F^{d+2}} T_0^{4+d} \ ,
\ee
where $G_{4+d}$ is the fundamental (4+d dimensional)
gravitational constant and $g_0$
is the density of states. The appropriately redshifted energy
density today is
\be
\label{ed}
G_{4+d}\rho_4 = \frac{g_0}{M_F^{d+2}} T_0^{4+d}
\left(\frac{\sigma_0}{\sigma_1}\right)^{d+4}
\left(\frac{a_1}{a_2}\right)^3
\left(\frac{a_2}{a_3}\right)^4
\left(\frac{a_3}{a_4}\right)^3 \ ,
\ee
where we have used $\rho \sigma^{d+4} = \mathrm{constant}$ for $t_0<t<t_1$
and correspondingly redshifted powers in the matter and
radiation dominated eras. On the other hand, the magnitude of the appropriately
redshifted curvature term today is
\be \label{curv}
\frac{1}{a_4^2} = \frac{1}{a_0^{2}}
\left(\frac{a_0}{a_1}\right)^2
\left(\frac{a_1}{a_2}\right)^2
\left(\frac{a_2}{a_3}\right)^2
\left(\frac{a_3}{a_4}\right)^2 \ .
\ee

Note that unlike inflationary models where the solutions of the entropy 
problem and flatness problem are closely linked, the large volume of the extra dimensions 
implied by our reformulation of the entropy problem 
does not imply a small value for the initial curvature $a_0^{-2}$, nor consequently the
present curvature $a_4^{-2}$.  The flatness problem therefore requires further consideration.

If we require that the evolution of the extra dimensions not disturb
the usual thermal history of our universe we need
\begin{eqnarray}
T_4 & = & 10^{-3} \mbox{ eV} \nonumber \\
T_3 & = & 10 \mbox{ eV} \\
T_2 & \gsim & 1 \mbox{ MeV} \nonumber \ .
\end{eqnarray}
Now, if we assume that the total entropy is conserved for $t_0<t<t_1$, i.e.
$S_0=S_1$, we obtain
\be
\left(\frac{\sigma_0}{\sigma_1}\right)=\left(\frac{g_1}{g_0}\right)^{1/(3+d)} 
\left(\frac{T_1}{T_0}\right) \ .
\ee
A similar consideration for $t_1<t<t_*$ yields
\be
\label{a121}
\left(\frac{a_2}{a_1}\right)^3=\left(\frac{a_*}{a_1}\right)^3 =
\left(\frac{g_1}{g_*}\right) \left(\frac{T_1}{T_*} \right)^{d+3} \ ,
\ee
We used $b_2 = b_*= b_1$ where $b_1$, the curvature scale at late times 
(including currently),
characterizes the low-energy mass of KK modes
\be
M_{KK} \sim b_1^{-1} \ ,
\ee
with $\beta$ constrained by relation (\ref{Vextra})
\be
e^\beta =
\left(\frac{M_{Pl}}{M_F}\right)^2
\left(\frac{M_{KK}}{M_F}\right)^d \ .
\ee
On the other hand, during the matter dominated era $t_1<t<t_2$,
when the extra dimensions are frozen, the relationship between the
scale factor and the age of
the universe is:
\be
\label{a122}
\left(\frac{a_2}{a_1}\right) =
\left(\frac{t_2}{t_1}\right)^{2/3} \ ,
\ee
with
\begin{eqnarray}
t_2  & \equiv & \tau_{KK} \sim \frac{M_{Pl}^2}{M_{KK}^3} \nonumber \\
t_1 & \sim & (G_{4+d} \rho_1)^{-1/2} = \left[\frac{g_0}{M_F^{d+2}} T_0^{4+d}
\left(\frac{\sigma_0}{\sigma_1}\right)^{d+4}
\right]^{-1/2} \ .
\end{eqnarray}
Eliminating $T_1$ using (\ref{a121}) and (\ref{a122}), we can
express all the relevant quantities in terms of dynamical
variables. Thus,
\be
\frac{\sigma_0}{\sigma_1} =
\frac{1}{ g_*}\left(\frac{M_{KK}^{6}M_F^{d+2}}{T_*^{d+3} M_{Pl}^4 T_0}\right) \ ,
\ee
\be
\left(\frac{a_2}{a_0}\right)^3 = \left(\frac{g_0}{g_*}\right)
\left( \frac{b_0}{b_1} \right)^{d}
\left(\frac{T_0}{T_*}\right)^{d+3} \ ,
\ee
\be
\left(\frac{a_2}{a_1}\right)^3 =
\left(\frac{M_{Pl}^4}{M_{KK}^{6}}\right) \frac{g_0}{M_F^{(d+2)}}
T_0^{(4+d)}
\left(\frac{\sigma_0}{\sigma_1}\right)^{(d+4)} \ .
\ee

We can now calculate the ratio of the two terms relevant for the
flatness problem: 
\be
\label{flatness} 
\frac{G_{4+d}\rho_4}{(1/a_4^2)} \sim 
\left(\frac{M_{KK}}{T_2}\right)^{14/3+8d/9} 
\left(\frac{M_{KK}^2 T_3 T_4}{M_{Pl}^4}\right)
\left(b_0 T_0\right)^{2d/3}
\left(a_0 T_0\right)^2 \frac{g_0^{2/3}g_*^{2d/9} }{g_2^{2(d+3)/9}}
\ee
It is not difficult to choose generic values of parameters which
yield this ratio significantly greater than one. However, by requiring a consistent
dynamical evolution (for example $a_0 < a_1 < a_2$, values of $T_0$
and $T_1$ not much greater than the fundamental quantum gravity
scale $M_F$ etc.) we considerably narrow the choice. One possible choice
(neglecting the contribution from the density of states) 
$a_0^{-1} \sim M_F \sim TeV$, $M_{KK} \sim 27 M_F$, $d=7$, $T_0 \sim 4 M_F$, 
$b_0 \sim 5 \times 10^5 M_F^{-1}$ and $T_2 \sim 130\mathrm{MeV}$ gives the numerical
value of this ratio to be about $10$ which reproduces the current flatness of the universe. 
Note that although some fine tuning of $b_0$ is present, the situation is
much better than in the ordinary 3-dim case where we needed to tune $a_0$ to 
about $30$  orders of magnitude.  

More heuristically, the explanation of cosmological flatness in this picture 
is the enormous injection of entropy into the brane by the combination of the
collapse of  the extra-dimensions to their final value, and the subsequent
decay of the KK modes in the bulk into modes on the brane.

We should briefly comment on the possibility that in the context of extra dimensions
the flatness problem may not be present {\it ab initio}. 
This is because the  structure of spacetime  may be a solution of 
the highly non-linear string equations of motion with some configuration of sources 
(eg. D-branes).  Suppose that the structure of the total space-time  is
\be
\Sigma_{4+d_1+d_2} = R \times \M_{\rm FRW}^3 \times \M^{d1}_{\rm extra1} \times
\M^{d2}_{\rm extra2} \ ,
\ee
where $\M^{d1}_{\rm extra1}$ is CHM. 
The source configuration may then require some specific $\M^{d2}_{\rm extra1}$, such 
as a hypersphere.  
The zero-global-curvature of $\M_{\rm FRW}^3$ may then be merely a consistency
condition of the solution, and we would then need  only explain the absence
of local inhomogeneities.

\subsection{The Homogeneity Problem}
\label{Homogeneity}
Now consider the homogeneity problem. We will see that
the process of entropy injection from extra dimensions embeds a huge number
of initially uncorrelated layers into the brane universe. Thus, the homogeneity of the 
brane universe
today may be greatly enhanced over that expected from the standard cosmology.

We assume that in the formation of the universe, there exists some correlation scale 
$\xi\simeq M_F^{-1}$,
on which fluctuations in all quantities (e.g. $\rho$) are correlated, 
but above which all fluctuations are independent.  We assume further that the 
fluctuations on this
scale are ${\cal O}(1)$.  In the absence of a complete underlying theory of the 
formation of the universe,
we offer no proof of this assumption.  
Other equally reasonable assumptions could undoubtedly be made.

Consider then a primordial fluctuation in
homogeneity $\delta \rho_0 / \rho_0 $. The magnitude of
this fluctuation, when evolved to
the present day, is suppressed by a huge
number $\sqrt{n}$, where $n$ is the number of appropriately  redshifted
fundamental volumes (of radii $1/M_{F}$) contained in the
horizon volume of the 3-space at some late time $t_4$ (which we take
to be the time of  last scattering, when $T_4 \sim 1$eV.)
\be
n=\frac{ e^\beta b_0^d (t_4/t_3)^3
(t_3/t_2)^3 (t_2/t_1)^3 (t_1/t_0)^3 t_0^3}{ \xi^{3+d}
(\sigma_1/ \sigma_0)^{d+3} (a_2/a_1)^3 (a_3/a_2)^3 (a_4/a_3)^3}\ .
\ee
In addition,
the primordial fluctuations contain a factor which grows in time.
In the radiation and matter dominated eras, the growth factors are
$t_f/t_{in}$ and $(t_f/t_{in})^{2/3}$ respectively, where the
subscripts ``in" and ``f" stand for ``initial" and ``final"
\cite{Bardeen}. Thus, the fluctuations at the horizon scale are:
\be
\left( \frac{\delta \rho}{\rho}
\right)_{Hor(t_4)} \sim \frac{1}{\sqrt{n}}
\left(\frac{t_4}{t_3}\right)^{2/3}
\left(\frac{t_3}{t_2}\right)\left(\frac{t_2}{t_1}\right)^{2/3}
\left(\frac{t_1}{t_0} \right)^{k} \left( \frac{\delta \rho}{\rho}
\right)_{Hor(t_0)} \ ,
\ee
where we have assumed that the
universe was effectively matter dominated for $t_1<t<t_2$,
when the radius of extra dimensions was frozen and most of the KK
excitations were massive. We leave the coefficient $k$
undetermined for now.

Using the relation between the scale factor and time for
$t_0<t<t_1$ $(t_1/t_0) = (a_1/a_0)^m$, where the coefficient $m$ is
also undetermined for now, we obtain:
\begin{eqnarray}
\left( \frac{\delta
\rho}{\rho} \right)_{Hor(t_4)} & \sim & e^{-\beta /2}
(b_0 M_F)^{-d/2} (t_0 M_F)^{-3/2} (\xi M_F)^{(3+d)/2}
\left(\frac{\sigma_1}{\sigma_0}\right)^{(d+3)/2} \nonumber \\
& & \times
\left(\frac{a_4}{a_3}\right)^{1/4}
\left(\frac{a_3}{a_2}\right)^{1/2}
\left(\frac{a_2}{a_1}\right)^{1/4}
\left(\frac{a_1}{a_0}\right)^{m(k-3/2)}
\left( \frac{\delta \rho}{\rho}\right)_{Hor(t_0)} \ .
\end{eqnarray}

The values for the unknown coefficients $m$ and $k$, in the the
ordinary 3-dim universe are $m=3/2$, $k=2/3$ for the matter
dominated and $m=2$, $k=1$ for the radiation dominated universe.
In the presence of the extra dimensions these numbers are
different but we assume that $m \geq 1$ and $k \leq 1$ in order not
to violate the causality. Thus, in the most conservative case,
$m=k=1$, we have:
\begin{eqnarray}
\label{homogeneityeqn}
& & \hspace{-.6cm} \left( \frac{\delta \rho}{\rho} \right)_{Hor(t_4)}  \sim 
\left(\frac{T_2}{M_F}\right)^{\frac{d^2}{3}+\frac{17d}{9}+\frac{19}{6}}
\left(\frac{M_F}{M_{KK}}\right)^{\frac{d^2}{12} +\frac{95d}{36}+ \frac{31}{6} }
\left(\frac{M_{Pl}}{M_{F}}\right)^{d+2}
 \left(\frac{T_0}{M_F}\right)^{\frac{d+3}{3}} \times \nonumber \\
& \times & \left(\frac{M_F^2}{T_3 T_4}\right)^{\frac{1}{4}}
\left(b_0 M_F\right)^{-\frac{2d}{3}}
\left(t_0 M_F\right)^{-\frac{3}{2}} (\xi M_F)^{\frac{3+d}{2}} \ 
\frac{g_0^{\frac{1}{12}}g_2^{\frac{1}{36}(24+17d+3d^2)}}{
 g_*^{\frac{1}{36}(8d+3d^2)} }
\end{eqnarray}

Unlike the flatness of the universe, 
it is much easier to explain its homogeneity without any fine tuning. 
For example, 
neglecting the contribution from the density of states,
if $d=7$, then $T_2 \sim 100$MeV with
\be
T_0 \sim b_0^{-1} \sim t_0^{-1} \sim \frac{M_{KK}}{10} \sim M_F \sim \mbox{TeV} \ ,
\ee
gives
\be
\left( \frac{\delta \rho}{\rho}\right)_{Hor(t_4)} = 10^{-8}
\left( \frac{\delta \rho}{\rho} \right)_{Hor(t_0)} \ .
\ee
which reproduces the current cosmological homogeneity  
if $\left( \frac{\delta \rho}{\rho} \right)_{Hor(t_0)}$ is of order
one, i.e. if initially the energy density distribution in the 
universe was peaked around the reasonable value of 
$T_0^{4+d}$. Note that if we use the same numbers which
explain flatness then we obtain an even more homogeneous universe
(with the $10^{-8}$ above replaced by $10^{-40}$).
Any initial inhomogeneities are thus smoothed out beyond detection purely by
the very large statistical averaging inherent in the collapse of
entropy onto the brane.  

It is worth pointing out that this relationship between the horizon and
flatness problems is opposite to that in inflation.
There
it is the horizon problem which is more 
difficult to solve, and the required expansion yields an extremely flat universe. This
is the reason that open universes from inflation \cite{Bucher:1995gb} require more 
than one period of inflation. In 
our case, one could imagine an open universe resulting from the dynamics we have 
described, with the horizon problem still remaining easily solved,
as can be seen from the absence of $\alpha$ in Eq. (\ref{homogeneityeqn})

For comparison, a similar calculation for the ordinary FRW case,
without the extra dimensions, and in which the fundamental volume
has radius $M_{\Pl}^{-1}$, gives
\be
\left( \frac{\delta \rho}{\rho} \right)_{Hor(t_4)} \sim
\frac{M_{\Pl}^{1/2}}{T_3^{1/4}T_4^{1/4}} \sim 10^{14} \left(
\frac{\delta \rho}{\rho} \right)_{Hor(t_0)} \ .
\ee

\section{Conclusions}
\label{Conclusions}
We have examined the problems of standard cosmology in a class
of brane-world models in which 
both the ordinary $3$-dimensional universe and 
the extra dimensions are  compact hyperbolic manifolds. 
In this context,
many of the problems of the standard cosmology are addressed in a new way.
First, the evolution of the extra-dimensional
space at early cosmic times can inject a huge
entropy onto the Standard-Model-supporting brane,
greatly enhancing the entropy inside the effective 3-dimensional horizon.
This allows us to re-express the large dimensionless number $S=10^{88}$ 
as an exponential of the ratio of the linear size of the extra-dimensions
of spacetime to their curvature scale -- a rather smaller dimensionless number.  
Although, unlike in inflation, 
we have offered no dynamical solution of the cosmological entropy problem
(the existence of the large dimensionless number $10^{88}$), we have reformulated
it in a way in which it may be less numerically daunting, and which
suggests a path for an alternative solution -- a dynamical understanding
of topology change in the quantum creation of the universe.
Interestingly, the necessary initial entropy of the full space need not exceed the 
holographic limit. 
Second, injection of the large initial entropy onto the brane from the extra dimensions 
results in a very
homogeneous brane universe today. 
 Finally, for reasonable parameters of the model (with a 
mild fine tuning in the extra space curvature scale), the curvature
of the 3-manifold is small today, and so the flatness  of the universe can be understood.
Thus, the evolution of the extra dimensional space in these models can result in
a low-energy universe, as seen from our brane, which is flat, full, and homogeneous.
Moreover, within  this framework, the recent
solutions to the hierarchy problem can be readily realized. 
We have offered no detailed fundamental model of the dynamics of the 
spacetime during the period before the extra dimensions are frozen, 
nor have we offered any calculation of the spectrum of primordial
fluctuations that would arise  in such a model. We suggest that
the appropriate dynamical models could  and should be found, 
and that sources for fluctuations do exist, at least in the dynamics
of the brane. This, and many other outstanding questions are the
subject of ongoing and future investigations.

We would like to thank Nemanja Kaloper for numerous invaluable
conversations and suggestions during the early stages of this project;
Andrei Linde for comments and discussions on an earlier version of this paper, and 
on the entropy problem and its solution 
in inflation;
also J. Alexander, R. Brandenberger, S. Carroll, G. Gibbons,
C. Gordon, J. Ratcliffe, W. Thurston,
T. Vachaspati and J. Weeks for many discussions and answers to
(for us) difficult questions.
The work of GDS and DS was supported by a DOE grant to the particle astrophysics
group at CWRU.  GDS  received additional support from an NSF CAREER award.
MT thanks the DOE and Syracuse University for support.

\end{document}